\documentclass[aps,showpacs,preprint,preprintnumber,nofootinbib,amsmath,amssymb,
ascmac,bm,12pt]{revtex4}
\usepackage{bm}
\usepackage[dvipdfmx]{graphicx}
\usepackage[dvips]{color} 
\usepackage{ascmac}

\begin{document}
\def\quote{\color{blue}}

\vspace*{0.5cm}
\title{
Charged rotating Kaluza-Klein multi-black holes and \\ 
multi-black strings 
in five-dimensional Einstein-Maxwell theory 
}
\bigskip
\author{
${}^{1,2}$Ken Matsuno\footnote{E-mail: matsuno@sci.osaka-cu.ac.jp},
${}^{1}$Hideki Ishihara\footnote{E-mail: ishihara@sci.osaka-cu.ac.jp},
${}^{3}$Masashi Kimura\footnote{E-mail: mkimura@yukawa.kyoto-u.ac.jp}
and
${}^{1}$Takamitsu Tatsuoka\footnote{E-mail: tatsuoka@sci.osaka-cu.ac.jp}
\bigskip
\bigskip
}
\affiliation{
${}^{1}$Department of Mathematics and Physics, Osaka City University, Sumiyoshi, 
Osaka 558-8585, Japan
\\
${}^{2}$Interdisciplinary Faculty of Science and Engineering, Shimane
University, Shimane 690-8504, Japan
\\
${}^{3}$Yukawa Institute for Theoretical Physics, Kyoto University, Kyoto 606-8502, 
Japan
\bigskip
\bigskip
}
\begin{abstract}
We construct exact solutions, 
which represent regular charged rotating Kaluza-Klein multi-black holes 
in the five-dimensional pure Einstein-Maxwell theory.  
Quantization conditions between the mass, the angular momentum, and charges appear 
from the regularity condition of horizon. 
We also obtain multi-black string solutions by taking some limits in the solutions.  
We extend the black hole solutions 
to the five-dimensional Einstein-Maxwell-Chern-Simons theory with an arbitrary 
Chern-Simons coupling constant. 
\end{abstract}

\preprint{OCU-PHYS-372}
\preprint{AP-GR-101}
\preprint{YITP-12-71}


\date{\today}
\maketitle

\section{introduction}
Higher-dimensional black holes in asymptotically Kaluza-Klein spacetimes 
would play an important role in understanding basic properties of 
fundamental theories,  
since our real observable world is a macroscopically four-dimensional spacetime, 
then extra dimensions must be compactified.  
Exact solutions of the Kaluza-Klein black holes are constructed explicitly, 
and their properties are studied. 
For example, 
five-dimensional squashed Kaluza-Klein black hole solutions  
\cite{Dobiasch:1981vh, Gibbons:1985ac, Gauntlett:2002nw, 
Gaiotto:2005gf, Ishihara:2005dp, Wang:2006nw, 
Yazadjiev:2006iv, Nakagawa:2008rm, Tomizawa:2008hw, Tomizawa:2008rh, Stelea:2008tt,  
Tomizawa:2008qr, Gal'tsov:2008sh, Matsuno:2008fn, Bena:2009ev, Tomizawa:2010xq, 
Mizoguchi:2011zj, Chen:2010ih, Stelea:2011fj, Nedkova:2011hx, Nedkova:2011aa, 
Tatsuoka:2011tx, Mizoguchi:2012vg}  
behave as fully five-dimensional black holes in the vicinity of 
the squashed S$^3$ horizon, 
while they asymptote to four-dimensional flat spacetimes 
with a twisted S$^1$ as a compactified extra dimension. 
Then we can regard these squashed Kaluza-Klein black hole solutions 
as models of realistic higher-dimensional black holes.

Recently, 
we have investigated extremely rotating regular vacuum multi-black holes 
in the five-dimensional asymptotically Kaluza-Klein 
spacetimes \cite{IPUC-85-1,Matsuno:2012hf}.  
We have shown that each black hole has a smooth horizon with the topology of 
the lens space in addition to the S$^3$. 
The compactness of the extra dimension plays an essential role for 
existence of multi-black holes in vacuum, 
and mass of each black hole is quantized by the size of the compactified 
extra dimension.

In the present paper, 
we generalize these results to charged rotating Kaluza-Klein multi-black hole 
solutions in the five-dimensional Einstein-Maxwell 
theory.   
Until now, to our knowledge, 
any exact charged rotating black hole solutions have not been found 
in the five-dimensional pure Einstein-Maxwell theory. 
In contrast, the five-dimensional Einstein-Maxwell-Chern-Simons theory, 
which is suggested by the supergravity, 
admits regular charged rotating 
black hole solutions 
in closed form 
both in asymptotically flat spacetimes  
\cite{Breckenridge:1996is,Cvetic:1996xz,Gauntlett:2002nw,Herdeiro:2002ft,
Herdeiro:2003un,Ortin:2004af,Chong:2005hr,Wu:2007gg}  
and in asymptotically Kaluza-Klein spacetimes  
\cite{Gauntlett:2002nw,Gaiotto:2005gf,Maeda:2006hd,Nakagawa:2008rm,
Tomizawa:2008hw,Tomizawa:2008rh,Stelea:2008tt,Tomizawa:2008qr,Matsuno:2008fn,
Gal'tsov:2008sh,Mizoguchi:2011zj,Mizoguchi:2012vg}.   
The Chern-Simons term is necessary for the solutions. 
Then, 
the solutions presented in this article are 
the first example of exact solutions that represent 
charged rotating multi-black holes in the five-dimensional 
pure Einstein-Maxwell theory.  

This paper is organized as follows.  
We present explicit forms of solutions in Sec.\ref{solution}, 
and investigate asymptotic structures of the solutions, 
the regularity at the horizons, and the conserved charges in Sec.\ref{Properties}.  
We show in Sec.\ref{limit} that  some known multi-black hole solutions are obtained 
by taking a limit in our solutions, 
and we also show in Sec.\ref{multiblackstrings} that multi-black string solutions 
are obtained by another limit. 
In Appendices, 
we generalize our solutions to the solutions of non-degenerate horizon case and 
solutions to the five-dimensional Einstein-Maxwell-Chern-Simons theory 
with an arbitrary value of the Chern-Simons coupling constant.

\section{solutions}\label{solution}
We consider charged rotating multi-black hole solutions in the 
five-dimensional Einstein-Maxwell theory with the action
\begin{align}
	S = \frac{1}{16\pi} \int d^5 x \sqrt{-g} 
		\left( R - F_{\mu\nu} F^{\mu\nu} \right) .
\end{align}  
We assume that the metric and the Maxwell field are written as 
\begin{align}
 ds^2 &= - H^{-2} dt^2 + H^2 (dx^2+dy^2+dz^2) 
		+ \left[ \alpha \left( H^{-1} -1 \right) dt + \frac{L}{2} d\psi 
		+ \beta \bm \omega \right] ^2 ,
\label{MET1} \\
	A_\mu dx^\mu &= \gamma H^{-1}  dt + \delta \bm \omega , 
\label{MAX1}
\end{align}
where 
\begin{align}
 H &= 1 + \sum_i \frac{m_i}{|\bm r - \bm r _i|} 
\label{FUNCH} 
\end{align}
is the harmonic function on the three-dimensional Euclid space 
with point sources located at $\bm r = \bm r_i := (x_i, y_i, z_i)$.   
The 1-form $\bm \omega $, 
which satisfies 
\begin{align}
	\nabla \times \bm \omega = \nabla H ,
\end{align}
has the explicit form 
\begin{align}
	\bm \omega = \sum_i m_i \frac{z-z_i}{|\bm r - \bm r _i|}
				\frac{(x-x_i)dy-(y-y_i)dx}{(x-x_i)^2+(y-y_i)^2} . 
\label{one-form}
\end{align}
In the expressions \eqref{MET1}-\eqref{one-form}, $m_i$ and $L$ are constants, 
and $\alpha, \beta, \gamma, \delta$ are parameters to be fixed. 
As will be shown later, $\bm r = \bm r_i$ are black hole horizons.

The five-dimensional Einstein equations  
require that the parameters $\alpha ,~\beta ,~\gamma ,~\delta $ should satisfy 
\begin{align}
 &2 \alpha ^2 - \beta ^2 +  4 \gamma ^2  -2 =0,
 \label{condeq1} \\
 &\alpha ^2 - 2 \beta ^2 - 4 \delta ^2 +2 =0,
 \label{condeq2}
\end{align}
and the Maxwell equations require
\begin{align}
\alpha  \gamma - \beta  \delta = 0 .
\label{condeq3}
\end{align}
From \eqref{condeq1}, \eqref{condeq2}, and \eqref{condeq3} we find  
two possible cases:
\begin{align}\label{cond1}
\alpha^2=\beta^2 , \quad \gamma^2 = \delta^2= \frac{2-\alpha^2}{4}, 
\end{align}
or 
\begin{align}\label{cond2}
\alpha^2= \frac{4}{3}\delta^2  , \quad \beta^2 =\frac{4}{3} \gamma^2 = 1-\alpha^2, 
\end{align}
where sign of the parameters should keep \eqref{condeq3}.

First, we assume $\beta\neq 0$, where 
the solutions describe multi-black holes. 
After that, we consider $\beta=0$ case, where the solutions describe 
multi-black strings.

\section{Charged Rotating Multi-Black Holes}\label{Properties}
We begin with the case $\beta\neq 0$.  In this case, the solutions describe 
charged rotating multi-black holes in the Einstein-Maxwell system.

\subsection{Asymptotic structure }

First, we see the asymptotic behavior of the metric \eqref{MET1}. 
In the limit $r\to \infty$, the metric behaves as 
\begin{align}
	ds^2 \to - dt ^2 +  dr^2 + r^2 d\Omega _{\rm S ^2}^2 
	+ \frac{ L ^2}{4} \left( 
		d\psi + \frac{2\beta}{L}\sum_i m_i \cos \theta d \phi \right) ^2 ,  
\label{asympt_metric}
\end{align}
where $d\Omega _{\rm S ^2}^2=d\theta^2+\sin^2\theta d\phi^2$ is the metric of 
the unit two-sphere. 
Then, the spacetime with the metric \eqref{MET1} is asymptotically 
locally flat, i.e., 
the metric asymptotes to a twisted constant S$^1$ fiber bundle over 
the four-dimensional Minkowski spacetime, 
and the spatial infinity has the structure of an S$^1$ bundle over an S$^2$. 
We see that the size of a twisted S$^1$ fiber as an extra dimension takes 
the constant value $L$ everywhere.

Next, we inspect the apparent singularities $\bm r = \bm r_i$ 
of the metric \eqref{MET1}. 
The absence of naked singularity 
on and outside the surfaces $\bm r = \bm r_i$ requires  
the parameters $m_i$ should be 
\begin{align}\label{PARAREGS}
	m _i > 0 . 
\end{align}
For simplicity, we restrict ourselves to the cases of two-black holes, 
i.e., $m _1\neq 0, m _2\neq 0$, otherwise $m _i = 0 $.
Without loss of generality, we can put the locations of two point sources as 
$\bm r_1 = (0, 0, 0)$ and $\bm r_2 = (0, 0, a)$, 
where the constant $a$ denotes the separation between two black holes.

In this case, the metric is  
\begin{align}\label{MET2}
	ds^2 = - H^{-2} dt^2 + H^2 
	\left( dr^2 + r^2 d\Omega _{\rm S ^2} ^2 \right) 
		+ \left[ \alpha \left( H^{-1} -1 \right) dt + \frac{L}{2} d\psi 
		+ \beta \bm \omega \right] ^2 ,
\end{align}
where  $H$ and $\bm \omega$ are given by 
\begin{align}
	H &= 1+ \frac{m_1}{r} 
		+ \frac{m_2}{\sqrt{r^2 + a^2 - 2 a r \cos \theta}} ,
\label{harmonics_2}
\\
	\bm \omega &= \left( 
		m_1 \cos \theta + m_2 \frac{r \cos \theta - a }
		{\sqrt{r^2 + a^2 - 2 a r \cos \theta}} \right) d\phi,
\label{form_2}
\end{align}
respectively. 
The coordinates run the ranges of 
$-\infty < t < \infty , ~0 < r < \infty,~0 \leq \theta \leq\pi,~ 0 \leq \phi \leq 2\pi $, and $0 \leq \psi \leq 4\pi $.

In the coordinate system $(t, r, \theta , \phi , \psi )$,
the metric \eqref{MET2} diverges at the locations of two point sources, i.e., 
$\bm r = \bm r_1 ~(r=0)$ and $\bm r = \bm r_2 ~(r=a, \theta=0)$. 
In order to remove apparent divergences 
at $r = 0$,
we introduce new coordinates $(v, \psi' )$ such that 
\begin{align}
	dv &= dt + H^2 dr + W d\theta ,
\\
	d\psi'
	&= d\psi - \frac{2 \alpha }{L}\left(dt + H dr + V d\theta 
		\right) , 
\end{align}
where the functions $W$ and $V$ are given by 
\begin{align}
	W \left(r, \theta\right)
		&= \int dr \frac{\partial}{\partial \theta}\left( H^2 \right) ,
\label{FUNCW}
\\
	V \left(r, \theta\right)&= \int dr \frac{\partial}{\partial \theta} H , 
\label{FUNCV}
\end{align}
respectively. 
Then, the metric \eqref{MET2} takes the form of 
\begin{align}
	ds^2 = & - H^{-2} \left(dv - W d\theta \right)^2 
		+ 2 dr \left(dv - W d\theta \right)
		+ H^2 r^2 d\Omega _{\rm S ^2} ^2
\notag \\
		& + \left[ 
			\alpha H^{-1} dv + \beta \bm \omega 
			+ \alpha \left( V - H^{-1}W \right) d\theta + \frac{L}{2} d\psi'
		\right]^2 . 
\label{MET3}
\end{align}

In the neighborhood of $r = 0$, 
the metric \eqref{MET3} behaves as 
\begin{align}
	ds^2 \simeq & \frac{(\alpha ^2 -1) r^2}{m_1 ^2} dv^2 + 2 dv dr 
		+ m_1 ^2 \left[ d\Omega _{\rm S ^2} ^2 
	+ \beta ^2 \left(\frac{L}{2 \beta m_1}d\psi'' 
		+ \cos \theta d\phi \right)^2 \right] 
\notag \\
		& + 2 r \left[ 
			\frac{2 m_1 m_2 \sin \theta }{a^2} dr d\theta 
			+ \alpha \beta \left( dv + \frac{3 m_1 m_2 \sin \theta }{2 a^2} r d\theta \right) 
			\left( \frac{L}{2 \beta m_1}d\psi'' + \cos \theta d\phi \right) 
	\right]  
\notag \\
	& + {\mathcal O}(r^3) ,
\label{near_horison_metric}
\end{align}
where we have used 
\begin{align}
	d\psi'' = d\psi' - \frac{2 \beta }{L} m _2 d\phi . 
\end{align}
If the factor $2 \beta m_1/L$ is a natural number, say $n_1$, 
the induced metric on the three-dimensional spatial cross section of 
$r = 0$ with 
a $t=\mbox{\it const.}$ surface 
is  
\begin{align}\label{INDMET}
	ds^2 |_{r=0} 
	= \frac{n_1^2 L^2}{4 \beta ^2} \left[ d\Omega _{\rm S ^2} ^2
     + \beta ^2 \left( \frac{d\psi''}{n_1} + \cos \theta d\phi \right)^2 \right] . 
\end{align}
That is, 
the $r=0$ surface admits the smooth metric of the lens space 
$L(n_1;1)=$ S$^3/\mathbb{Z}_{n_1}$.

In this case, from \eqref{near_horison_metric} 
we see that $r=0$ is a null surface where 
the metric \eqref{MET3} is regular and each component is 
an analytic function of $r$.
Therefore, the metric \eqref{MET3} gives analytic extension 
across the surface $r = 0$. 
By the same discussion, we see that 
the metric \eqref{MET2} also admits analytic extension across 
the surface $\bm r = \bm r _2$ 
if $2 \beta m_2/L$ is 
equal to  
a natural number $n_2$.

We also see that $\eta=\partial_v$ is a Killing vector field 
that becomes null at $r=0$. 
Furthermore, $\eta$ is hypersurface orthogonal to the surface $r=0$, i.e., 
$\eta_{ \mu} dx^\mu = g_{vr} dr =dr $ there. 
These mean that the null hypersurface $r=0$ is a Killing horizon. 
Similarly,  $\bm r = \bm r _2$ is also a Killing horizon. 
Hence, we can see that 
the solutions \eqref{MET2} with \eqref{harmonics_2} and \eqref{form_2} describe 
Kaluza-Klein multi-black holes, 
which have smooth Killing horizons without singularity on and outside 
the black hole horizons. 
Since the  $\phi$-$\psi$ part of the metric is positive definite, 
it is clear that no closed timelike curve exists. 
Near each horizon limit, the metric \eqref{MET2} 
approaches the $L(n_i;1)$ bundle 
over the AdS$_2$ space at the horizon  \cite{Reall:2002bh,Kunduri:2007vf}.

We can generalize the discussion 
to the metric \eqref{MET1}.   
The metrics \eqref{MET1} describe multi-black holes of   
the lens spaces $L(n_i;1)$ topology   
if the parameters $m_i$ are quantized as 
\begin{align}
	m_i =\frac{L}{2 \beta}n_i ,
\end{align} 
where $n_i$ are natural numbers.
The area of the horizons are given by 
\begin{align}
	\mathcal A|_{\bm r = \bm r _i} = \frac{n_i^2 L^3}{\beta ^2} \mathcal A_{\rm S ^3} , 
\end{align}
where $\mathcal A_{\rm S ^3}$ denotes the area of a unit S$^3$.

\subsection{Mass and Charges}

We define the Komar mass  $M$ associated with the timelike Killing vector 
$\xi_{(t)} = \partial / \partial t$, and obtain as 
\begin{align}
	M &= \frac{-3}{32\pi}\int_\infty dS_{\mu\nu}\nabla^\mu \xi_{(t)}^\nu
		= \frac{3 L \sum_i m_i}{4 \pi} \mathcal A_{\rm S ^3} .
\label{MASS}
\end{align}
We also obtain the angular momentum $J^{\psi}$ 
associated with the spacelike Killing vector 
$\xi_{(\psi)} = \partial / \partial \psi$ 
as
\begin{align}
	J^{\psi} &= \frac{1}{16\pi}\int_\infty dS_{\mu\nu}
		\nabla^\mu \xi_{(\psi)}^\nu
	= \frac{\alpha L ^2 \sum_i m_i}{8 \pi} \mathcal A_{\rm S ^3} .
\label{ANGMOM}
\end{align}
We see that the spacetime \eqref{MET1} is rotating along 
the extra dimension.

We can obtain the electric charge $Q_i$ and the magnetic flux $\Psi _i$ 
\begin{align}
	Q_i &= \frac{1}{4 \pi} \int_{\Sigma^3} {}^* \bm F 
		= \frac{\gamma L m_i}{\pi} \mathcal A_{\rm S ^3} ,
\label{elecharge} \\
	\Psi _i &= \frac{1}{4 \pi} \int_{\Sigma^2 } \bm F
		= - \delta m_i ,
\label{magcharge} 
\end{align}
respectively, 
where $\Sigma^3$ denotes a closed surface on a time slice 
surrounding each black hole, 
and $\Sigma^2$ denotes a closed surface surrounding each black hole 
on the base space.

In these calculations, 
we see that $m_i$ characterize the mass of black holes, 
and $\alpha, \gamma, \delta$ are related with the angular momentum, 
the electric charge, the magnetic flux, respectively. The parameter $L$ 
is related to the size of extra dimension at infinity as seen before. 
The parameter $\beta$ controls the twist of extra dimension, and gives 
the unit of quantization of mass in cooperation with $L$.  
If none of these parameters vanishes, the solutions describe 
electrically and magnetically charged rotating Kaluza-Klein multi-black holes 
to the Einstein-Maxwell equations.

From conditions \eqref{condeq1} and \eqref{condeq2}, we have
\begin{align}\label{condeqwa}
	\frac{1}{4} \left( \alpha ^2 +\beta ^2 \right) +\gamma ^2 +\delta ^2 -1 = 0 . 
\end{align}
Substituting the Komar mass \eqref{MASS}, the angular momentum \eqref{ANGMOM}, 
the electric charge \eqref{elecharge}, and the magnetic flux \eqref{magcharge} 
into \eqref{condeqwa}, 
we obtain the extremal condition as 
\begin{align}
\label{extremalcond}
	\left( \frac{4 J^\psi}{L} \right)^2 + Q^2 
	+ \left( 2 \pi L \Psi \right)^2 + \left( \frac{\pi n L^2}{2} \right)^2 
	= \left( \frac{4 M}{3} \right)^2 , 
\end{align}
where $Q = \sum _i Q_i ,~ \Psi = \sum _i \Psi _i$, and $n = \sum _i n_i$. 
We should note that the size of extra dimension $L$ cannot be infinitely 
large under the mass $M$ is finite. The Kaluza-Klein structure is a critical 
property for the present solutions. 
We can rewrite the conditions 
\eqref{cond1} and \eqref{cond2} for possible two cases 
in terms of $M ,~ J^\psi ,~ Q ,~ \Psi$ and $L$. 
In the first case \eqref{cond1} we have 
\begin{align}\label{extremalcond1}
	Q^2 = \left( 2 \pi L \Psi \right)^2, 
	\quad 
	\left( \frac{4 J^\psi}{L} \right)^2 
	=\frac12 \left( \frac{4 M}{3} \right)^2 - Q^2
	= \left( \frac{\pi n L^2}{2} \right)^2 . 
\end{align}
The angular momentum and the mass square minus the electric charge square are 
quantized by the size of extra dimension. 
In the second case \eqref{cond2} we have 
\begin{align}\label{extremalcond2}
	\left( \frac{4 J^\psi}{L} \right)^2 = \frac13 \left( 2 \pi L \Psi \right)^2 ,
	\quad 
	\frac{1}{3} Q^2 
	=\frac14 \left( \frac{4 M}{3} \right)^2 - \left( \frac{4 J^\psi}{L} \right)^2 
	= \left( \frac{\pi n L^2}{2} \right)^2 . 
\end{align}
The electric charge and the mass square minus the angular momentum square are 
quantized in this case.  
If $\Psi=0$, one of $Q$ or $J^\psi$ vanishes. 
It is clear that non-vanishing magnetic flux is a key for 
the charged rotating Kaluza-Klein multi-black hole solutions.

With respect to the timelike Killing vector 
$\xi_{(t)}$, we define 
the ergosurfaces where the Killing vector becomes null, i.e., 
\begin{align}\label{ERGOEQ}
	g_{tt} = -H^{-2} + \alpha ^2 \left( H^{-1} - 1 \right)^2 = 0 . 
\end{align}
Since $g_{tt}$ is a continuous function on the spacetime outside the horizons 
for the range of parameters \eqref{PARAREGS}, and  
$g_{tt}(\bm r = \bm r_i) = \alpha ^2 > 0$ and $g_{tt}(\infty) = -1 < 0$, 
then there always exist ergoregions 
around the black hole horizons. 
The topology of the ergosurface depends on the location of 
black holes \cite{Matsuno:2008fn}.

In the same manner, we can construct charged rotating single Kaluza-Klein black hole 
solutions with non-degenerate horizons to the Einstein-Maxwell equations. 
This case is discussed briefly in Appendix \ref{NONDEGBH}.
In Appendix \ref{EMCSBH},  
we also generalize our solutions to the solutions 
in the five-dimensional Einstein-Maxwell-Chern-Simons theory 
with an arbitrary value of the Chern-Simons coupling constant.

\section{Black holes with $\Psi _i = 0$ }\label{limit} 
We consider the limiting solutions with $\Psi _i = 0$. 
There are two subcases: $ \delta=\gamma=0,~ \alpha ^2 = \beta ^2 = 2$, and 
$\delta=\alpha=0,~ \beta^2 = \frac{4}{3} \gamma^2 = 1$.

In the first case, 
$Q_i = \Psi _i = 0$ then the Maxwell field \eqref{MAX1} vanishes. 
Then the metric \eqref{MET1} coincides with the Kaluza-Klein vacuum 
multi-black holes \cite{IPUC-85-1,Matsuno:2012hf}:
\begin{align}
 ds^2 &= - H^{-2} dt^2 + H^2 (dx^2+dy^2+dz^2) 
  + 2\left[(H^{-1}-1)dt +\frac{L}{2\sqrt{2}}d\psi \pm \bm \omega \right]^2 . 
\end{align}
In the second case, 
$J^\psi = \Psi _i = 0$ 
then the metric \eqref{MET1} and the Maxwell field \eqref{MAX1} reduce to  
\begin{align}
 ds^2 &= - H^{-2} dt^2 + H^2 (dx^2+dy^2+dz^2) 
	+ \left( \frac{L}{2} d \psi + \bm \omega \right) ^2 ,
 \\
 A_\mu dx^\mu &= \pm \frac{\sqrt 3}{2} H^{-1}  dt , 
\end{align}
which describe 
charged static Kaluza-Klein multi-black holes with a twisted constant S$^1$ 
\cite{Gauntlett:2002nw,Ishihara:2006iv}.

\section{Multi-black strings}\label{multiblackstrings} 
Here, we consider the case $\beta = 0$. 
In this case, the fiber-bundle structure reduces to the direct product 
of the S$^1$ fiber as the extra-dimension and base space. 
Then, the metrics \eqref{MET1} 
describe multi-black strings.

We have two subcases. 
In the first case, $\beta=\alpha=0$ and $\gamma^2 = \delta^2 = 1/2$,  
the metric \eqref{MET1} and the Maxwell field \eqref{MAX1} become 
\begin{align}
 ds^2 &= - H^{-2} dt^2 + H^2 (dx^2+dy^2+dz^2) + \frac{L^2}{4} d\psi ^2 ,
 \label{met01}
 \\
 A_\mu dx^\mu &= \pm \frac{1}{\sqrt 2} \left( H^{-1}  dt + \bm \omega \right) , 
 \label{max01}
\end{align}
which describe dyonically charged 
static multi-black strings.

In the second case, $\beta = \gamma= 0$ and $\alpha^2=\frac43\delta^2 =1$,  
the metric \eqref{MET1} and the Maxwell field \eqref{MAX1} reduce to  
\begin{align}
 ds^2 &= - H^{-2} dt^2 + H^2 (dx^2+dy^2+dz^2)
	+ \left( (H^{-1}-1)  dt + \frac{L}{2}d\psi \right) ^2 ,
\label{magstringmet}  \\
 A_\mu dx^\mu &= \pm \frac{\sqrt 3}{2} \bm \omega , 
 \label{magstringmax}
\end{align}
which describe magnetically charged boosted multi-black strings.

In the single black string case, 
the solution \eqref{met01} and \eqref{max01} was obtained 
in the context of the ten-dimensional type-IIA string 
theory in \cite{Sheinblatt:1997nt}, and
the solution \eqref{magstringmet} and \eqref{magstringmax} was obtained 
in the context of the five-dimensional Einstein-Maxwell-dilaton 
theory in \cite{Emparan:2004wy, Yazadjiev:2006wr}.  
Analytic extensions across the horizons for \eqref{met01} and \eqref{magstringmet}
are given by \eqref{MET3} with $\beta = 0$.

\section{Summary} \label{Discussions}
We have constructed exact charged rotating Kaluza-Klein multi-black hole solutions 
in the five-dimensional pure Einstein-Maxwell theory. 
The metric asymptotes to the effectively four-dimensional spacetime at infinity, 
and the size of the compactified extra dimension 
takes the constant value everywhere.  
We have shown that 
each black hole has a smooth horizon and its topology is the three-dimensional 
sphere or lens space $L(n_i;1)$ with an arbitrary $n_i$. 
The positions of black holes on the three-dimensional flat base space are 
free parameters.

The solutions are characterized by 
the size of extra dimension, 
the mass, the angular momentum along the extra circular dimension, 
the electric charge, and the magnetic flux. 
These quantities are related with three conditions 
which come from the Einstein-Maxwell equations. 
Regularity of horizons requires that some of these quantities are quantized 
by the size of extra dimension, $L$. 
Then, the minimum size of black hole which is comparable to $L$ exists. 
By this reason, we cannot obtain asymptotically flat solutions from the present 
solutions 
by taking the limit $L \to \infty$ keeping the black hole mass finite. 
This is consistent with the fact that 
any exact charged rotating black hole solutions in asymptotically flat spacetimes 
have not been found yet 
in the five-dimensional pure Einstein-Maxwell theory.\footnote{
There are some attempts to obtain charged rotating asymptotically flat black hole solutions  
numerically \cite{Kunz:2005nm, Kunz:2005ei, Kunz:2006yp, Kunz:2006xk}
and 
perturbatively \cite{Aliev:2005npa, Aliev:2006yk, NavarroLerida:2007ez, Allahverdizadeh:2010xx}.}      
The Kaluza-Klein spacetime structure with a compact extra dimension 
plays a crucial role in constructions of exact charged rotating black hole solutions 
in the five-dimensional pure Einstein-Maxwell theory.

We have also obtained multi-black string solutions by taking limits. 
Furthermore, we can easily generalize the solutions in the five-dimensional 
Einstein-Maxwell-Chern-Simons theory 
with an arbitrary value of the Chern-Simons coupling constant 
(see Appendix \ref{EMCSBH}).

\section*{Acknowledgments}
This work is supported by the Grant-in-Aid for Scientific Research No.19540305. 
M.K. is supported by the JSPS Grant-in-Aid for Scientific Research No.11J02182.

\appendix
\section{Charged rotating Kaluza-Klein black holes with non-degenerate horizons}
\label{NONDEGBH}
We consider the metric and the Maxwell field of 
the charged rotating Kaluza-Klein black holes with non-degenerate horizons 
in five dimensions as 
\begin{align}
	ds^2 = &- \left( 1 -\frac{2m}{R} +\frac{q^2}{R^2} \right) dt^2 
		+ \left( 1 -\frac{2m}{R} +\frac{q^2}{R^2} \right)^{-1} dR^2 
		+ R^2 d\Omega_{\rm S ^2} ^2
	\notag \\
	   &+ \left( \frac{L}{2} d\psi -\alpha \frac{q}{R} dt 
		+ \beta q \cos \theta d\phi \right) ^2 ,
\label{NDMET1} \\
	A_\mu dx^\mu =& -\gamma \frac{q}{R} dt + \delta q \cos \theta d\phi . 
\label{NDMAX1}
\end{align}
The Einstein equations require 
the conditions \eqref{condeq1} and \eqref{condeq2} between the parameters 
$\alpha ,~\beta ,~\gamma ,~\delta $ 
and the Maxwell equations require \eqref{condeq3}, 
as same as the multi-black hole case.

For the absence of naked singularity $q \leq m$ then we have 
\begin{align}\label{normalcond}
	\left( \frac{4 J^\psi}{L} \right)^2 + Q^2 
	+ \left( 2 \pi L \Psi \right)^2 + \left( \frac{\pi n L^2}{2} \right)^2 
	\leq \left( \frac{4 M}{3} \right)^2 ,
\end{align}
instead of \eqref{extremalcond}.
If $m = q$, by the use of $r=R-m$, 
we recover the single black hole case of the metric \eqref{MET1} 
and the Maxwell field \eqref{MAX1} 
with $m_1=m$ and $m_i=0~ (i\geq 2)$.  

The metric \eqref{NDMET1} and the Maxwell field \eqref{NDMAX1} is discussed 
as a special solution, where the size of the extra dimension is constant, 
to the Einstein-Maxwell-Chern-Simons equations 
in Ref.~\cite{Nakagawa:2008rm}. The parameters 
in their solution are different from the present solution.

\section{Charged rotating black holes in Einstein-Maxwell-Chern-Simons system}
\label{EMCSBH}
We consider 
the five-dimensional Einstein-Maxwell-Chern-Simons theory with the action
\begin{align}\label{actionEMCS}
	S = \frac{1}{16\pi} \int d^5 x 
\left[ 	
	\sqrt{-g} \left( R - F_{\mu\nu} F^{\mu\nu} \right)
	- \frac{2 \lambda }{3 \sqrt 3} \epsilon^{\mu\nu\rho\sigma\zeta}
	A_\mu F_{\nu \rho } F_{\sigma \zeta }
\right] ,
\end{align}  
where $\lambda$ is the Chern-Simons coupling constant  
\cite{Gauntlett:1998fz, Kunz:2005ei, Kunz:2006yp, Kunz:2006xk, Aliev:2008bh, Allahverdizadeh:2010xx}.   
For vanishing $\lambda$, pure Einstein-Maxwell theory is recovered, 
and $\lambda=1$ is suggested by the minimal supergravity.

We assume the same forms of the metric and the Maxwell field \eqref{NDMET1} and 
\eqref{NDMAX1} 
for single black holes or \eqref{MET1} and \eqref{MAX1} for multi-black holes. 
Since the Chern-Simons term is free from the metric, the Einstein equations 
require the same conditions \eqref{condeq1} and \eqref{condeq2}. 
On the other hand, the Maxwell equations modified by the Chern-Simons term 
require 
\begin{align}\label{condeq3cs}
 3 (\alpha  \gamma - \beta  \delta ) -4 \sqrt{3} \gamma  \delta \lambda = 0 .
\end{align}
If the parameters $\alpha, \beta, \gamma, \delta$ satisfy 
\eqref{condeq1}, \eqref{condeq2}, and \eqref{condeq3cs},  
the metric and the Maxwell field is a solution 
which represents charged rotating Kaluza-Klein black holes 
in the Einstein-Maxwell-Chern-Simons theory with arbitrary $\lambda$. 
In the $\lambda=1$ case, the single black hole solution 
coincides with a special case of solutions obtained 
in \cite{Nakagawa:2008rm, Mizoguchi:2011zj}.



\end{document}